# INTERETS ET LIMITES DE L'APPROCHE MOLECULAIRE POUR ABORDER LA BIOGEOGRAPHIE ET LA SPECIATION : L'EXEMPLE DE QUELQUES MAMMIFERES D'AFRIQUE TROPICALE

par

Sophie QUÉROUIL

Dans cette étude, nous nous sommes proposés de tester les hypothèses biogéographiques formulées pour la faune tropicale (refuges, barrières fluviales, gradients environnementaux...), en appliquant l'approche moléculaire la plus employée, le séquençage d'ADN mitochondrial, à quelques taxons de Mammifères africains (Insectivora, Rodentia, Primates). Nous avons d'abord tenté d'obtenir une phylogénie moléculaire d'espèces potentiellement intéressantes pour la phylogéographie, dans le but de vérifier leur monophylie et de calibrer une horloge moléculaire. Puis, nous avons recherché et comparé les schémas phylogéographiques de quatre espèces de petits Mammifères et d'une super-espèce de Primates, dans le but d'en tirer des conclusions biogéographiques. Enfin, nous avons évalué les processus évolutifs potentiellement impliqués.

Nos résultats phylogénétiques confirment que l'histoire des gènes n'est pas forcément celle des taxons et qu'il est important de prendre en compte plusieurs sources d'information indépendantes. Les analyses phylogéographiques révèlent des scénarios différents pour chacun des modèles, ce qui peut refléter soit des distributions initiales différentes, soit une réponse différentielle aux mêmes événements selon les taxons. Ces scénarios présentent une certaine concordance avec les régions fauniques définies pour les forêts d'Afrique centrale, mais suggèrent que les événements de divergence intra-spécifique seraient beaucoup plus anciens que les derniers cycles glaciaires. Nos résultats sont en accord avec la théorie des refuges et confirment le rôle de barrière joué par les principales rivières.

# Powers and limitation of the molecular approach for the study of biogeography and speciation : examples in tropical African mammals

We attempted to test biogeographic hypotheses proposed for the evolution of tropical faunas (e.g. refuge, riverine barrier, and environmental gradient theories) by applying the most widely used molecular approach, i.e. mitochondrial DNA sequencing, to selected African mammalian taxa (Insectivora, Rodentia and Primates). First, we constructed a molecular phylogeny of taxa we intended to use for phylogeographic studies, in order to ascertain their monophyly and to calibrate a molecular clock for divergence time estimates. Second, we analysed and compared the phylogeographic patterns of four forest-dwelling small mammal species and one primate super-species. Third, we evaluated the evolutionary processes potentially involved in the speciation of cercopithecine Primates, by testing their geographic mode of speciation and reconstructing evolutionary scenarios of some life-history traits.

Our phylogenetic results confirm that gene history is not necessarily the same as organism history. Thus, mitochondrial DNA should be studied in combination with other independent data, such as nuclear genes, morphology, ecology and behaviour. The obtained phylogeographic patterns all differ one from the other, which may be explained by differences in initial distributions, or by different responses to the same events. However, individual patterns present a certain degree of consistency with the faunal areas defined for the central African forest. They indicate a role of Plio-Pleistocene vicariance events in the intra-specific diversification of small mammals and suggest that genetic divergence would be much older than the last glacial cycles. In the case of cercopithecine Primates, speciation would have been predominantly allopatric and driven by Miocene and Pliocene vicariance events. Taken together, our results give support to the refuge hypothesis, without excluding the riverine barrier nor the paleogrographic hypotheses. They emphasize the role of paleo-ecological changes in generating diversity and that of the main riverine barriers in shaping the present distribution of that diversity.

## Introduction

Pour comprendre l'évolution du monde vivant, il est nécessaire de connaître la façon dont les organismes se diversifient dans le temps et dans l'espace. C'est l'objectif de la biogéographie, discipline qui étudie la distribution des taxons à l'échelle des peuplements, repère les caractères communs à ces distributions, et cherche à identifier les processus environnementaux, biologiques et historiques qui les ont façonnés (MYERS & GILLER, 1988). Une approche plus récente, baptisée phylogéographie (AVISE *et al.*, 1987), consiste à étudier l'évolution d'une lignée à la fois, en intégrant une dimension phylogénétique. La comparaison de scénarios obtenus pour différentes lignées, ou phylogéographie comparée, tend à rejoindre les objectifs de la biogéographie.

Les études biogéographiques et phylogéographiques réalisées dans diverses régions du monde ont permis de proposer des hypothèses évolutives concernant la flore et la faune. La plupart des modèles évolutifs attribuent un rôle majeur à la fragmentation des aires de distribution (allopatrie). Plusieurs de ces modèles font intervenir les changements paléo-environnementaux liés aux mouvements tectoniques (théorie paléogéographique - Emsley, 1965) et aux oscillations climatiques (théorie des refuges - HAFFER, 1969, qui attribue un rôle majeur aux zones de stabilité environnementale, et théorie des perturbations - COLINVAUX, 1993, BUSH, 1994, qui considère au contraire les modifications de l'habitat dans les zones d'écotone). Un autre modèle évolutif met l'accent sur la notion de barrière physique à la dispersion (théorie des barrières fluviales - CAPPARELLA, 1991, AYRES & CLUTTON-BROCK, 1992). Des études récentes ont réhabilité le concept de diversification en présence de flux génique (sympatrie). Notamment, en ce qui concerne la faune, les variations locales de l'environnement floristique pourraient être suffisantes pour initier une diversification des taxons par sélection naturelle (théorie des gradients environnementaux - ENDLER, 1982).

La grande diversité de la faune des régions tropicales a suscité de nombreux travaux et hypothèses. Cependant, l'essentiel des travaux récents qui ont permis d'alimenter le débat sur les modes de spéciation en milieu tropical se sont focalisés sur l'Amérique du Sud (MORITZ *et al.*, 2000). La plupart des études réalisées en Afrique tropicale ne sont basées que sur la distribution actuelle des taxons et la reconnaissance de zones d'endémisme et de diversité élevée (e.g. CARCASSON, 1964 ; MOREAU, 1969 ; GRUBB, 1990 ; COLYN, 1991). Des travaux intégrant la notion de parenté entre assemblages faunistiques ont été menés par notre équipe sur la faune mammalienne d'Afrique centrale (DELEPORTE & COLYN, 1999 ; COLYN & DELEPORTE, 2002). Ils ont permis d'affiner le découpage en régions fauniques qui avait été proposé auparavant. Ils ont révélé une structure complexe, avec un emboîtement de régions et de sous-régions fauniques séparées par des zones de mélange des faunes appelées zones d'intergradation (Fig. 1). Cette structure met en évidence le rôle de barrière joué par les principales rivières. Elle pourrait s'expliquer par la théorie des refuges : dans chaque sous-région, il y aurait eu un refuge forestier où la faune forestière aurait survécu pendant les phases d'aridité et aurait évolué en allopatrie. Les études des variations paléo-climatiques et paléo-environnementales en Afrique tropicale viennent à l'appui de cette hypothèse. Elles suggèrent qu'il y aurait eu fragmentation de la forêt tropicale au cours des périodes glaciaires, froides et sèches, et expansion forestière durant les interglaciaires, chauds et humides (e.g. MALEY, 1996).

Ces travaux se heurtent au fait que la taxinomie et la distribution des espèces sont souvent mal connues. Il leur manque une dimension historique car ils ne prennent pas en compte les relations phylogénétiques entre taxons, ni leur distribution passée, d'autant plus difficile à connaître que les fossiles sont rares en forêt tropicale. Un moyen d'y circonvenir est d'utiliser l'approche moléculaire, qui permet de révéler l'histoire des taxons recelée par leur matériel génétique (e.g. AVISE, 2000). Cette approche génère des caractères supposés indépendants des conditions environnementales, et dont l'évolution peut avoir été suffisamment régulière pour qu'il soit possible de calibrer une "horloge moléculaire" et de dater les événements de divergence. La technique la plus prisée actuellement est celle du séquençage d'ADN mitochondrial, car elle permet d'obtenir rapidement un grand nombre de caractères variables, en s'affranchissant des problèmes de recombinaison et de polymorphisme individuel propres à l'ADN nucléaire.

Nous nous sommes proposés d'évaluer les schémas biogéographiques proposés pour les forêts de plaines africaines et les hypothèses évolutives sous-jacentes, en reconstituant l'histoire de différents taxons de Mammifères à partir de séquences d'ADN mitochondrial. Nous avons cherché quelles conclusions biogéographiques pouvaient être tirées de la comparaison des variations géographiques de la diversité génétique pour différents modèles. Nous avons également tenté de savoir si certaines des prédictions associées aux théories évolutives formulées pour la faune tropicale étaient vérifiées.

## Matériel et méthodes

La zone principale d'étude correspond à une région de l'Afrique centrale pour laquelle nous disposons d'un échantillonnage conséquent et qui semble présenter à elle seule toute la complexité qui peut être observée à plus large échelle (Fig. 1). Cependant, certaines questions ont été abordées à l'échelle de l'ensemble des forêts de plaine africaines, ainsi qu'à l'interface forêt-savane. Les modèles biologiques ont été choisis parmi trois Ordres de Mammifères : Insectivores, Rongeurs et Primates. Les Primates, moins bien représentés dans nos collections, ont été beaucoup plus étudiés que les petits Mammifères sur le plan biologique et écologique, ce qui nous a permis d'aborder plus en détail les mécanismes de la spéciation. Les méthodes utilisées pour obtenir des séquences d'ADN et reconstruire des phylogénies sont décrites ailleurs (QUEROUIL, 2001). Les principales méthodes de phylogéographie et de biogéographies utilisées seront mentionnées au fur et à mesure. L'essentiel des analyses est basé sur un fragment d'ADN mitochondrial codant pour l'ARN-r 16S (ci-après appelé haplotype).

## Résultats et discussion

**Choix des modèles**

Le choix des modèles a nécessité un travail de révision taxinomique et phylogénétique préliminaire. Les analyses phylogénétiques réalisées pour les musaraignes africaines (QUEROUIL *et al.*, 2001) et pour le genre de Rongeurs africains *Hylomyscus* ont permis de retenir deux espèces d'Insectivores (*Sylvisorex johnstoni* et *S. ollula*) et une espèce de Rongeurs (*Hylomyscus stella*) comme modèles pour les analyses phylogéographiques. Deux autres modèles ont été choisis parmi des taxons ne présentant pas de difficulté taxinomique particulière : *Stochomys longicaudatus*, une espèce de Rongeurs appartenant à un genre monospécifique, et *Cercopithecus cephus*, une super-espèce de Primates.

**Analyses phylogéographiques**

Les analyses phylogéographiques réalisées pour la super-espèce *C. cephus* mettent en garde contre les limites de l'utilisation de l'ADN mitochondrial pour reconstruire des phylogénies. Ils confirment que l'histoire des gènes n'est pas forcément celle des taxons (e.g. MOORE, 1995) et qu'il est important de prendre en compte plusieurs sources d'information indépendantes, telles que des gènes non liés sur la même molécule, la morphologie, l'écologie et le comportement (e.g. GRANDCOLAS *et al.*, 2001).

Les schémas phylogéographiques obtenus pour *S. johnstoni* et *S. longicaudatus* présentent une ségrégation géographique marquée des haplotypes qui serait due à des événements de vicariance anciens (Fig. 2). Les schémas obtenus pour *S. ollula* et *H. stella* font apparaître une expansion récente dans la zone d'étude (Fig. 2). Afin d'identifier les mécanismes impliqués, nous avons analysé la structuration géographique de la diversité génétique par une analyse des groupements emboîtés (TEMPLETON, 1998). La majorité des événements résulterait de la fragmentation passée de l'aire de distribution (vicariance). Il y aurait eu aussi plusieurs événements de dispersion et d'isolement par la distance.

Les analyses phylogéographiques ont révélé quatre scénarios phylogéographiques différents pour les quatre modèles de petits Mammifères retenus, ce qui tendrait à confirmer que "la concordance entre schémas phylogéographique est l'exception plutôt que la règle" (ZINK, 1996 ; TABERLET *et al.*, 1998). Contrairement à ce qui apparaît dans d'autres études comparatives portant sur les Rongeurs du bassin amazonien (PATTON *et al.*, 2000) ou sur des régions géographiques limitées d'Amérique du Nord et d'Australie (TABERLET, 1998), les schémas phylogéographiques obtenus ne sont concordants ni dans le temps ni dans l'espace, et ceci même pour des espèces phylogénétiquement proches. Ce résultat pourrait s'expliquer par des différences dans le cadre spatial et temporel de l'évolution des espèces. Le paramètre temporel ne semble pas prédominant car au moins trois des quatre espèces seraient apparues vers la fin du Pliocène ou le début du Pléistocène (c'est-à-dire à la charnière entre Terciaire et Quaternaire). En revanche, comme les distributions géographiques des espèces au cours du Pléistocène ne sont pas connues, il est possible que les différences phylogéographiques observées résultent, au moins en partie, de différences d'extension et de localisation des aires de distribution passées.

Une autre possibilité est que toutes les espèces aient bien été sympatriques et soumises à des événements similaires durant la majeure partie du Pléistocène, mais n'aient pas répondu de la même façon aux fluctuations paléo-environnementales. La variabilité des réponses aux changements paléo-environnementaux pourrait être liée à la survie différentielle des espèces dans différents refuges, à leur sensibilité aux barrières et aux pressions environnementales, leurs capacités de colonisation et leur pouvoir compétiteur, ainsi qu'à différents paramètres liés à la dynamique des populations (ZINK, 1996 ; TABERLET *et al.*, 1998). Toutes ces hypothèses font intervenir les caractéristiques écologiques et biologiques des espèces, or celles-ci sont peu connues dans le cas des modèles étudiés.

**Synthèse biogéographique**

Dans un premier temps, nous avons essayé de savoir si la diversité génétique intra-spécifique a été affectée par l'histoire supposée des aires biogéographiques. Chaque schéma phylogéographique a été comparé avec un arbre théorique des aires correspondant aux régions fauniques mises en évidence pour la faune mammalienne (Fig. 1). Il est apparut que les arbres obtenus étaient significativement différents de l'arbre théorique des aires pour trois des quatre espèces (pour la quatrième espèce, *H. Stella*, le non-rejet de l'hypothèse nulle résulterait d'une faible différentiation génétique). La ségrégation géographique des haplotypes ne suit donc pas exactement les régions fauniques ni les relations présumées entre aires. Cependant, certains groupements d'haplotypes coïncident avec les régions fauniques (Fig. 2). Nos résultats tendent à confirmer l'identité des régions Ouest Congo (à l'exception de la localité Odzala) et Sud Ogooué, tout en mettant en évidence des échanges entre ces deux régions. La localité Odzala semble avoir eu une histoire particulière, fonctionnant comme une zone de convergence pour *S. ollula* et *H. stella*, et comme une zone génétiquement isolée pour *S. johnstoni* et *S. longicaudatus*. La région supposée d'intergradation pourrait avoir été colonisée à partir de différentes régions pour trois des modèles, mais apparaît comme une région génétiquement distincte dans le cas de *S. johnstoni*. Les donnés sont insuffisantes pour confirmer l'hypothèse d'un corridor savanicole ayant existé pendant la majeure partie du Pléistocène, où se serait produit un mélange des faunes lors des phases d'expansion récente de la forêt (cf. DELEPORTE & COLYN, 1999).

Dans un second temps, nous avons essayé de reconstruire l'histoire biogéographique de la région à partir de nos données phylogéographiques. La plupart des méthodes classiques de biogéographie cladistique ne sont pas applicables pour un nombre aussi limité de modèles, aussi avons nous choisi la méthode de HOVENKAMP (1997) qui présente en outre l'avantage de ne pas nécessiter la prédéfinition des aires biogéographiques. Cette méthode, améliorée en tenant compte des temps de divergence estimés, nous a permis de conforter deux événements de vicariance, observés dans deux cladogrammes et dont les datations concordent (QUEROUIL *et al.*, 2002). Les résultats ne nous ont pas permis de proposer de nouveaux scénarios biogéographiques pour l'Afrique centrale, mais suggèrent que ce but pourrait être atteint en analysant un nombre beaucoup plus élevé de modèles. Ils remettent en cause la prédiction que, s'il existe une histoire commune des aires, elle devrait être reflétée par la majeure partie des peuplements (NELSON & PLATNICK, 1981).

**Ancienneté des événements de divergence**

Les principaux événements de divergence ont été datés après calibration d'une horloge moléculaire propre à chaque taxon, en se basant sur les datations proposées pour les fossiles. La diversité génétique intra-spécifique semble remonter au Pliocène pour au moins deux des modèles (*S. johnstoni* et *S. longicaudatus*), l'événement de divergence le plus ancien ayant été estimé à 3,2 +/- 0,6 millions d'années (MA). Les phénomènes à l'origine de la différenciation intra-spécifique seraient donc bien antérieurs aux derniers cycles glaciaires supposés avoir modelé les régions fauniques (contra GRUBB, 1990 ; COLYN, 1991), lesquels peuvent par contre expliquer la distribution géographique actuelle des lignées, en conjonction avec les conditions écologiques, géomorphologiques et climatiques (cf. ENDLER, 1982). Au niveau inter-spécifique, trois des espèces de petits Mammifères seraient apparues vers la fin du Pliocène, il y a au moins 2 MA (*S. johnstoni* et *S. longicaudatus*), ou au début du Pléistocène, il y a environ 1,8 MA (*S. ollula*). La quatrième espèce, *H. stella*, pourrait être apparue plus tard (quelque part entre 1,9 et 0,6 MA). De même, chez les Primates de la tribu des Cercopithecini, les événements de divergence entre espèces soeurs remonteraient pour la plupart au Pliocène.

Ces datations sont plus anciennes que celles qui ont été envisagées à partir des fossiles, mais sont en accord avec les datations moléculaires réalisées pour de nombreux taxons de Mammifères. En effet, près d'un tiers des événements de divergence intra-spécifique et deux-tiers des événements de divergence entre espèces soeurs seraient antérieurs au Pléistocène (Avise, 2000). Les événements de spéciation conduisant aux espèces actuelles seraient donc aussi anciens en Afrique tropicale que dans d'autres régions tropicales et tempérées.

**Recherche des processus évolutifs impliqués**

Nous avons tenté de savoir si certaines des prédictions associées aux théories évolutives qui ont été formulées pour la faune forestière tropicale étaient vérifiées. Nous avons notamment évalué l'importance de la vicariance et de l'allopatrie, principes sur lesquels reposent la plupart des méthodes biogéographiques et des théories évolutives, à l'aide d'un test du mode géographique de spéciation appliqué aux Primates de la tribu des Cercopithecini. Les résultats indiquent une prédominance de l'allopatrie et des événements de vicariance du Miocène et du Pliocène dans l'évolution de ce taxon. Ils sont en accord avec ce qui a été observé chez les petits Mammifères, chez qui la différentiation intra-spécifique résulterait essentiellement d'événements de vicariance ayant eu lieu au Plio-Pléistocène.

Nous avons aussi essayé de déterminer si les pressions environnementales pouvaient avoir favorisé la spéciation. Chez les Cercopithecini, chacune des espèces occupe une gamme d'habitats variés et les espèces soeurs occupent des habitats semblables, ce qui tend à minimiser le rôle de l'habitat comme cause initiale de la spéciation. Dans le cas des petits Mammifères, il ne semble pas que les caractéristiques de l'habitat aient induit une différentiation intra-spécifique car la diversité génétique n'est pas corrélée avec les assemblages floristiques.

Nous avons également abordé la question de la permanence des préférences écologiques au cours de l'évolution, qui est un postulat implicite à la théorie des refuges. Pour cela, nous avons construit une phylogénie des Cercopithecini en utilisant l'ensemble des données génétiques, morpho-anatomiques et écologiques disponibles, puis pisté l'évolution de différentes caractéristiques éco-éthologiques sur cette phylogénie. En dépit de la plasticité écologique et comportementale de certaines espèces, il y aurait eu un nombre limité de changements d'habitat et de mode de locomotion chez les Cercopithecini.

Dans l'ensemble, nos résultats tendent à exclure les théories faisant intervenir les pressions environnementales comme moteur principal de l'évolution, i.e. la théorie des gradients environnementaux et celle des perturbations (Tab. 1). Ils confortent surtout les prédictions associées à la théorie des refuges et à la théorie paléogéographique, sans exclure la théorie des barrières fluviales, dont l'effet pourrait s'être additionné à celui des fluctuations paléo-environnementales. La théorie paléogéographique pourrait être éliminée *a priori* en raison de la rareté des événements géologiques ayant affecté la zone d'étude (bien que la formation du rift Est Africain puisse avoir eu des répercussions morphogénétiques et climatiques en Afrique centrale). La théorie des refuges est la théorie le plus souvent invoquée pour expliquer la distribution actuelle de la faune dans les forêts de plaine africaines (e.g. CARCASSON, 1964 ; MOREAU, 1969 ; GRUBB, 1990 ; COLYN, 1991). Il est généralement admis que la forêt tropicale africaine a effectivement été fragmentée durant des périodes de forte aridité du Terciaire et surtout du Quaternaire (MALEY, 1996 ; LINDER, 2001). En revanche, la théorie des refuges est controversée dans le cas de l'Amérique du Sud, faute de preuve de la fragmentation de la forêt amazonienne (MORITZ *et al.*, 2000 ; PATTON *et al.*, 2000). Ainsi, alors que le principal moteur de l'évolution pourrait être d'origine paléo-climatique en Afrique tropicale, il serait probablement d'origine morphogénétique en Amérique tropicale.

## Conclusion et perspectives

Ces travaux peuvent être considérés comme une étude exploratoire permettant d'évaluer les intérêts et les limites des analyses phylogéographiques dans l'étude de l'évolution des faunes forestières d'Afrique tropicale, et de mettre en évidence des pistes qui mériteraient d'être explorées à l'avenir. Dans l'optique de tester des hypothèses biogéographiques d'ordre général, telles que l'influence de la latitude, l'impact de barrières biogéographiques prédéfinies ou l'importance du mode de dispersion, il apparaît nécessaire d'étudier un grand nombre de modèles diversifiés. Cependant, une analyse "en profondeur" privilégiant quelques modèles semble nécessaire pour évaluer les processus mis en jeu. Nous suggérons de multiplier les points de collecte en Afrique centrale et d'adopter une approche multidisciplinaire, de façon à déterminer avec précision le statut taxinomique et l'aire de distribution des espèces étudiées, et à pouvoir trier les espèces selon un filtre écologique. Cette perspective ne peut être envisagée que comme un travail d'équipe et sur le long terme.

Université de Rennes I, CNRS - U.M.R. 6552, Laboratoire Ethologie - Evolution - Ecologie, Station Biologique, 35380 Paimpont, France.
Adresse actuelle : Instituto do Mar (IMAR), Departamento de Oceanografia e Pescas, Cais Santa Cruz, 9901-862 Horta, Portugal. squerouil@dop.horta.uac.pt

## RÉFÉRENCES

AVISE, J.C. (2000).- *Phylogeography*. Harvard University Press, Cambridge, MA.

AVISE, J.C., ARNOLD, J., BALL, R.M., BERMINGHAM, E., LAMB, T., NEIGEL, J.E., REEB, C.A. & SAUNDERS, N.C. (1987).- Intraspecific phylogeography: The mitochondrial DNA bridge between population genetics and systematics. *Annu. Rev. Ecol. Syst.*, **18**, 489-522.

AYRES, J.M. & CLUTTON BROCK, T.H. (1992).- River boundaries and species range size in Amazonian primates. *Am. Nat.*, **140**, 531-537.

BUSH, M.B. (1994).- Amazonian speciation: a necessarily complex model. *J. Biogeogr.*, **21**, 5-17.

CAPPARELLA, A.P. (1991).- Neotropical avian diversity and riverine barriers. *Ibid.*, **20**, 307-316.

CARCASSON, R.H. (1964).- A preliminary survey of the zoogeography of African butterflies. *East Afr. Wildl. J.*, **2**, 122-157.

COLINVAUX, P. (1993)- Pleistocene biogeography and diversity in tropical forests of South America. *In* Biological Relationships Between Africa and South America. Goldblatt, P. (Ed.), New Haven, CT, Yale Univ. Press, 437-499.

COLYN, M. (1991).- L'importance géographique du bassin du fleuve Zaire pour la spéciation : le cas des Primates simiens. *Ann. Sci. Zool. Mus. R. Afr. Cent. Tervuren, Belgique*, **264**, 1-250.

COLYN, M. & DELEPORTE, P. (2002).- Biogeographic analysis of Central African forest guenons. *In* Multicolored monkeys: diversity and adaptation in the guenons of Africa. Glenn, M., Cords, M. (Eds), Kluwer Academic / Plenum.

DELEPORTE, P. & COLYN, M. (1999).- Biogéographie et dynamique de la biodiversité: application de la "PAE" aux forêts planitiaires d'Afrique centrale. *Biosystema*, **17**, 37-43.

EMSLEY, M.G. (1965).- Speciation in *Heliconius* (Lep., Nymphalidae): morphology and geographic distribution. *Zoologica (NY)*, **50**, 191-254.

ENDLER, J.A. (1982).- Pleistocene forest refuges: fact or fancy. *In* Biological diversification in the Tropics. Prance, G.T. (Ed.), Columbia University Press, New York, 641-657.

GRANDCOLAS, P., DELEPORTE, P., DESUTTER-GRANDCOLAS, L. & DAUGERON, C. (2001).- Phylogenetics and ecology: As many characters as possible should be included in the cladistic analysis. *Cladistics*, **17**, 104-110.

GRUBB, P. (1990).- Primate geography in the Afro-Tropical forest biome. *In* : Vertebrates in the Tropics. Peters, G., Hutterer, R. (Eds), Museum Alexander Koenig, Bonn, 187-214.

HAFFER, J. (1969).- Speciation in Amazonian forest birds. *Science*, **165**, 131-136.

HOVENKAMP, P. (1997).- Vicariance events, not areas, should be used in biogeographical analysis. *Cladistics*, **13**, 67-79.

LINDER, H.P. (2001).- Plant diversity and endemism in sub-Saharan tropical Africa. *J. Biogeog.*, **28**, 169-182.

MALEY, J. (1996).- The African rain forest - main characteristics of changes in vegetation and climate from the Upper Cretaceous to the Quaternary. *Proc. R. Soc. Edinburgh*, **104B**, 31-73.

MOORE, W.S. (1995).- Inferring phylogenies from mt-DNA variation: mitochondrial-gene versus nuclear-gene tree. *Evolution*, **49**, 718-726.

MOREAU, R.E. (1969).- Climatic changes and the distribution of forest vertebrates in west Africa. *J. Zool. Lond.*, **158**, 39-61.

MORITZ, C., PATTON, J.L., SCHNEIDER, C.J. & SMITH, T.B. (2000).- Diversification of rainforest faunas: An integrated molecular approach. *Annu. Rev. Ecol. Syst.*, **31**, 533-563.

MYERS, A.A. & GILLER, P.S. (1988).- *Analytical biogeography: an integrated approach to the study of animal and plant distributions*, Chapman & Hall, London.

NELSON, G. & PLATNICK, N. (1981).- *Systematics and biogeography. Cladistics and vicariance.* Columbia University Press, New York.

PATTON, J.L., DA SILVA, M.N.F. & MALCOLM, J.R. (2000).- Mammals of the Rio Jurua and the evolutionary and ecological diversification of Amazonia. *Bull. Am. Mus. Nat. Hist.*, **244**, 13-306.

QUEROUIL S. (2001).- *Intérêts et limites de l'approche moléculaire pour aborder la biogéographie et la spéciation : l'exemple de quelques Mammifères d'Afrique tropicale.* Thèse de doctorat, Université de Rennes 1, 310 p.

QUEROUIL S., HUTTERER R., BARRIERE P., COLYN M., PETERHANS J.C.K. & VERHEYEN E. (2001).- Phylogeny and evolution of African shrews (Mammalia : Soricidae) inferred from 16s rRNA sequences. *Mol. Phylogenet. Evol.*, **20 (2)**, 185-195.

QUÉROUIL S., COLYN M., VERHEYEN E. & DELEPORTE P. (2002).- De la phylogéographie comparée à la biogéographie : l'exemple de quelques Mammifères forestiers d'Afrique centrale. *Biosystema*, **20**, 131-140.

TABERLET, P. (1998).- Biodiversity at the intraspecific level: The comparative phylogeographic approach. *J. Biotech.*, **64**, 91-100.

TABERLET, P., FUMAGALLI, L., WUST SAUCY, A.G. & COSSON, J.F. (1998).- Comparative phylogeography and post-glacial colonization routes in Europe. *Mol. Ecol.*, **7 (4)**, 453-464.

TEMPLETON, A.R. (1998).- Nested clade analyses of phylogeographic data: testing hypotheses about gene flow and population history. *Mol. Ecol.*, **7**, 381-397.

ZINK, R.M. (1996).- Comparative phylogeography in North American birds. *Evolution*, **50**, 308-317.

Tableau 1 : Prédictions associées aux différentes théories de l'évolution. Caractères gras : prédiction vérifiée par nos résultats, soutenant la théorie concernée. Caractères soulignés : prédiction non vérifiée, fournissant un argument contre la théorie.

| Prédictions | Gradients | Perturbations | Barrières fluviales | Paléo-géographie | Refuges |
|---|---|---|---|---|---|
| mode géographique de spéciation | <u>sympatrie</u> | **allopatrie** | **allopatrie** | **allopatrie** | **allopatrie** |
| spéciation par différentiation écologique | <u>oui</u> | <u>oui</u> | **non** | **non** | **non** |
| ancienneté des divergences entre espèces | fin IV | IV | fin IV | **III-IV** | **III-IV** |
| concordance géographique entre taxons (et par rapport aux ...) | non | non | oui **(rivières)** | oui (événements géologiques) | oui **(régions fauniques)** |
| concordance temporelle entre taxons (et par rapport aux ...) | non | oui (périodes froides) | non | oui (événements géologiques) | oui (périodes arides) |
| phases de contraction / expansion des populations | non | **oui** | non | **oui** | **oui** |

**Légende des figures**

**Figure 1**. Carte des régions fauniques observées pour la faune mammalienne en Afrique centrale, d'après DELEPORTE & COLYN (1999). Les régions et sous-régions fauniques sont délimitées par différents niveaux de gris. Les régions appartenant à une même unité faunique sont regroupées par une ligne pointillée. Les zones hachurées représentent les régions d'intergradation, dans la limite du bloc forestier. Les principales localités de collecte sont indiquées sous forme abrégée et la zone principale de l'étude est encadrée en gras.

**Figure 2**. Mise en correspondance des groupements phylogénétiques obtenus pour les quatre modèles petits Mammifères avec les régions fauniques. Les pointillés rassemblent les localités dont les haplotypes se groupent lors des analyses phylogéographiques. Les flèches indiquent des événements probables de dispersion sur longue distance. Les surfaces grisées représentent des régions génétiquement homogènes, dont les limites hypothétiques sont adaptées des travaux de DELEPORTE & COLYN (1999).

**Figure legends**

**Figure 1**. Map of mammalian faunal areas in Central Africa, after DELEPORTE & COLYN (1999). The regions and sub-regions are shaded in grey, and the intergradation areas are dashed. The main collection sites are indicated by a three-letter code, and the main study area is delimited by a square box.

**Figure 2**. Comparison of haplotype clusters revealed by phylogeographic analyses of four small mammal species with the faunal areas. Dashed lines circle localities sharing closely related haplotypes. Arrows represent potential dispersal events. Shaded areas delineate genetically homogenous regions, which are hypothetically delimited according to DELEPORTE & COLYN (1999).

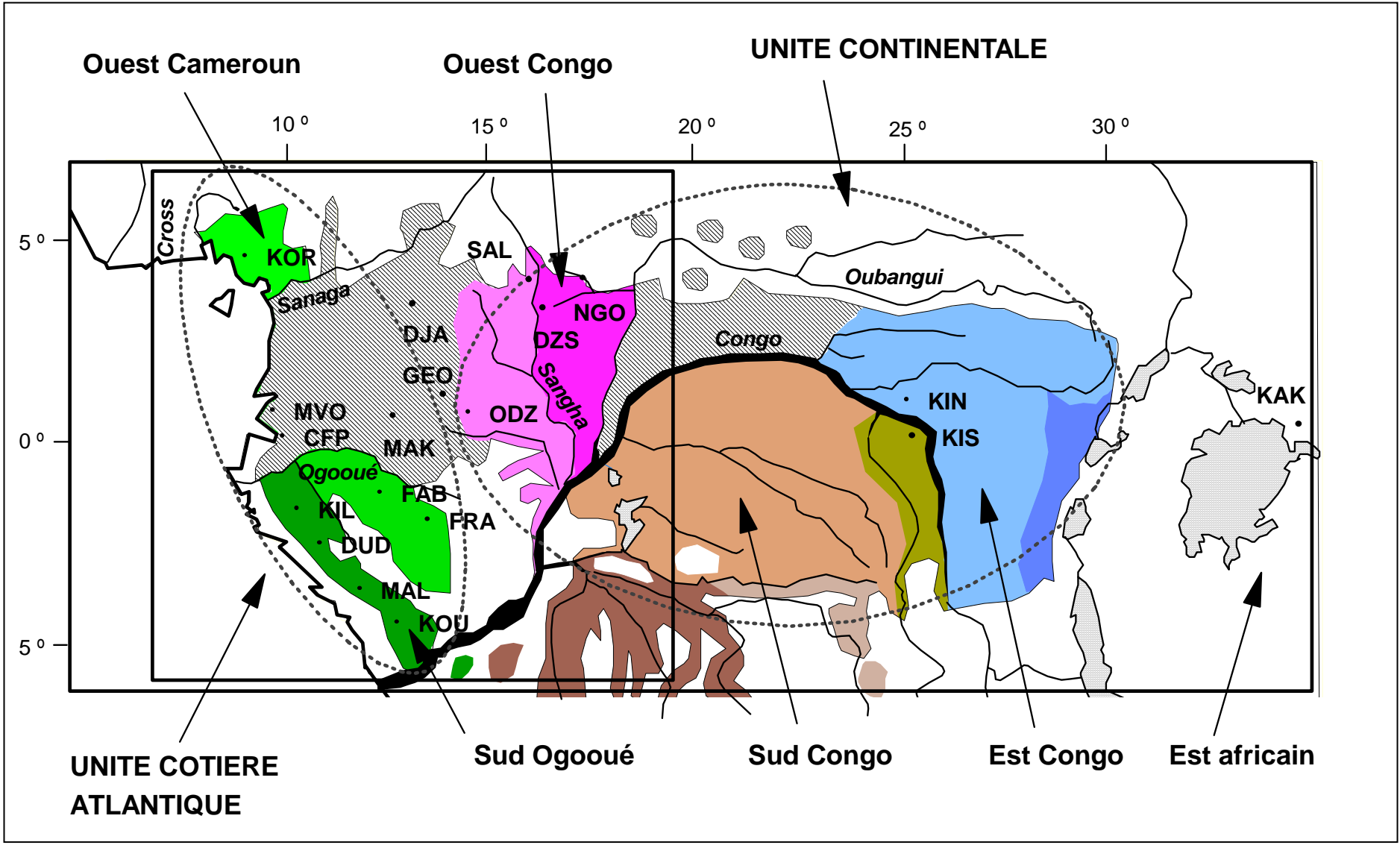

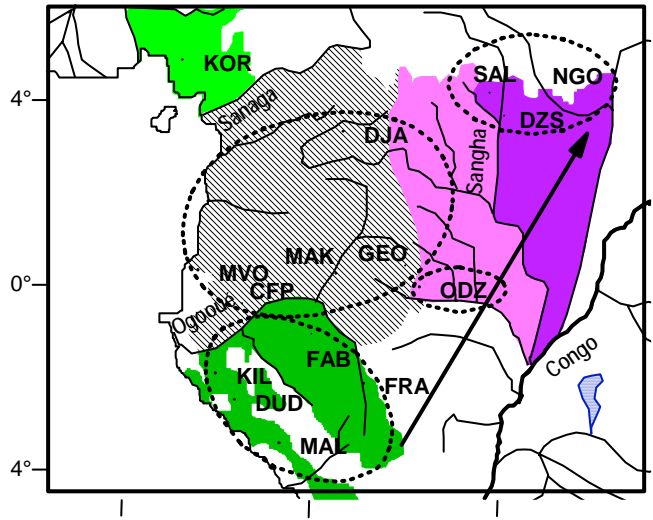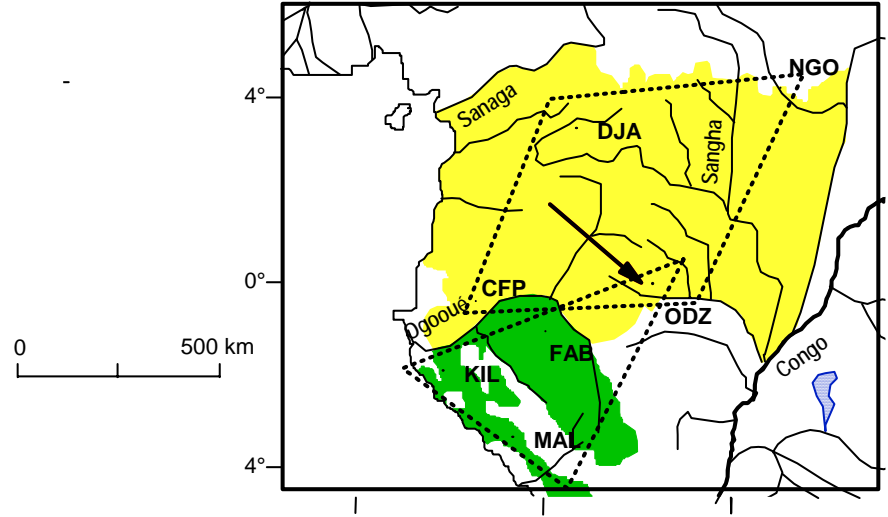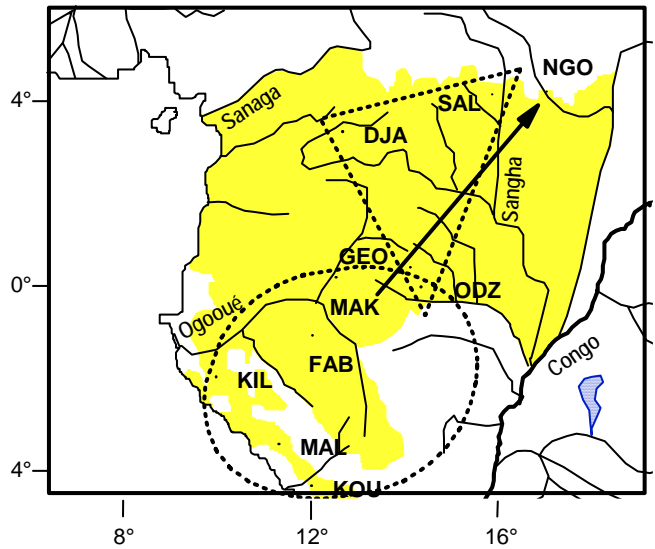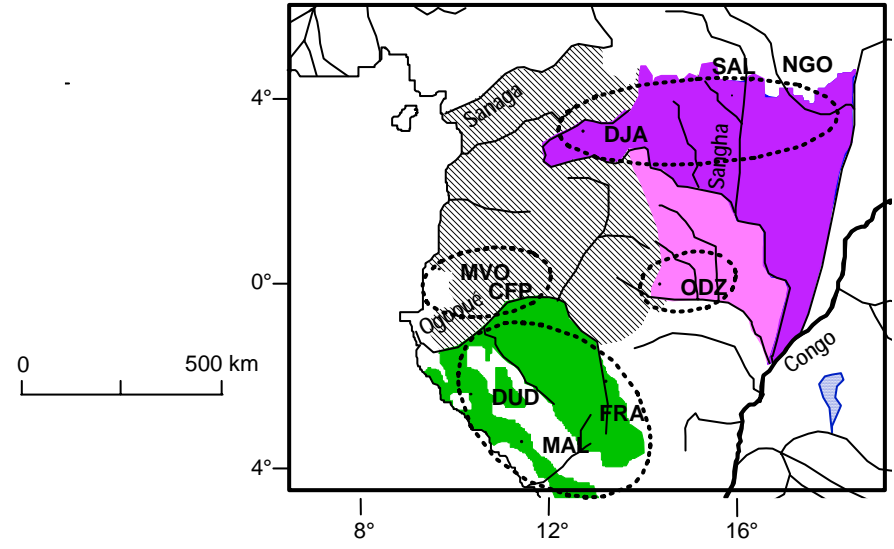
15